\documentclass[apl,reprint,amsmath]{revtex4-1}

\usepackage[dvips]{graphicx}
\usepackage{graphics}
\usepackage{amssymb}
\usepackage{amsmath}
\usepackage{epsfig}
\usepackage{latexsym}
\usepackage{rotating}
\usepackage{color}
\usepackage{ulem}

\renewcommand{\section}[1]{\textit{#1}--}

\begin{document}

\title{Low-loss flake-graphene saturable absorber mirror for laser mode-locking at sub-200-fs pulse duration}
\author{B. V. Cunning}
\affiliation{Queensland Micro- and Nanotechnology Centre, Griffith University, Nathan QLD 4111, Australia}
\author{C. L. Brown}
\affiliation{Queensland Micro- and Nanotechnology Centre, Griffith University, Nathan QLD 4111, Australia}
\author{D. Kielpinski}
\affiliation{Centre for Quantum Dynamics, Griffith University, Nathan QLD 4111, Australia}

\begin{abstract}
Saturable absorbers are a key component for mode-locking femtosecond lasers. Polymer films containing graphene flakes have recently been used in transmission as laser mode-lockers, but suffer from high nonsaturable loss, limiting their application in low-gain lasers. Here we present a saturable absorber mirror based on a film of pure graphene flakes. The device is used to mode lock an erbium-doped fiber laser, generating pulses with state-of-the-art, sub-200-fs duration. The laser characteristic indicate that the film exhibits low nonsaturable loss (13\% per pass) and large absorption modulation depth (45\% of low-power absorption).
\end{abstract}

\maketitle

Femtosecond laser pulses are essential tools for applications ranging from atomic clocks \cite{Udem-Hansch-optical-metrology-review} to medical imaging \cite{Boppart-Fujimoto-femtosecond-OCT}. Femtosecond lasers use saturable absorption to initiate pulse formation in a process called mode-locking. Short laser pulses with high peak intensity experience less loss in a saturable absorber than long, low-intensity pulses, so gain competition in the laser favors short pulses. While the saturable absorption effect can be obtained by various indirect means, the simplest and most robust femtosecond lasers employ physical optical elements that are engineered to exhibit the desired intensity-dependent absorption characteristic. The most widely used of these devices is the semiconductor saturable absorber mirror (SESAM), which uses heterostructure quantum wells as absorbers deposited on a reflective substrate \cite{Keller-compact-ultrafast-rev}. At high laser intensity, the small number of available conduction-band states is rapidly filled so that absorption ceases. More recently, it has been realized that nanoparticles can yield the desired saturable absorption, and devices based on single-walled carbon nanotubes (SW-CNTs) are now well established as saturable absorbers for femtosecond lasers \cite{Hasan-Ferrari-nanotube-graphene-optics-rev}.

Since the emergence of graphene as an optical material, a number of experiments have used graphene as a saturable absorber material to mode-lock lasers. Macroscopic single- or multi-layer graphene sheets are high-quality saturable absorbers \cite{Zhang-Loh-graphene-absorber-ultrafast, Martinez-Yamashita-exfoliated-graphene-MLL} and can exhibit loss at the few percent level in the saturated state, allowing their use for femtosecond pulse generation in bulk lasers \cite{Lee-Acosta-graphene-Er-glass, Cho-Rotermund-graphene-CrForsterite-100fs}. Devices using micron-scale graphene flakes dispersed in polymer also exhibit large saturable absorption \cite{Hasan-Ferrari-nanotube-graphene-optics-rev, Zhang-Loh-flake-graphene-ultrafast, Sun-Ferrari-graphene-Er-fiber, Martinez-Yamashita-optically-deposited-graphene-ML}. Suspensions of graphene flakes are reliably and straightforwardly prepared by a number of chemical processes \cite{De-Coleman-surfactant-graphene-flake-film, Lotya-Coleman-surfactant-stabilised-graphene-dispersion, Su-Li-electrochemical-exfoliation-graphene-flakes}, avoiding the specialized and/or probabilistic preparation procedures used for sheet graphene. A pure flake-graphene saturable absorber was used to mode-lock an Nd:YAG laser, achieving 4 ps pulse duration \cite{Tan-Tang-graphene-NdYAG-4ps}. In recent work, a saturable absorber based on graphene flakes in polymer achieved 175 fs pulse duration, a record for any graphene mode-locked fiber laser \cite{Popa-Ferrari-graphene-Er-fiber-175fs}. However, the loss of the absorber in that experiment was 70\%, while the change in loss due to saturation was 2\%, making it unsuitable for low-gain lasers. Moreover, exclusively transmissive geometries have been used for flake-graphene devices, whereas the most commonly used semiconductor saturable absorbers are reflective.

Here we present a low-loss flake-graphene saturable absorber mirror (FG-SAM) and demonstrate its ability to mode-lock a laser at 190 fs pulse duration. In contrast to previous flake-graphene devices for femtosecond pulse generation, the active layer of our FG-SAM is pure graphene, without a polymer matrix, and is used in a reflective geometry. Our FG-SAM was prepared by sonication of bulk graphite, creation of a thin film from the resulting graphene flakes, and transfer of this film onto a mirror substrate. This process was performed by a method analagous to that previously reported \cite{De-Coleman-surfactant-graphene-flake-film}. Gold mirrors (250 nm Au thickness) were prepared from atomically flat silicon using an Ar$^{+}$ sputterer. The stable graphene dispersions were prepared by bath sonicating 0.5 g flake graphite (Sigma-Aldrich) in 50 mL of 1\% solution of Triton-X 100 for 18 hours followed by overnight sedimentation. The resulting supernatant solution was then centrifuged twice at 1000 $g$ for 1 hour retaining the supernatant liquid in each case. 7.5 mL of this centrifuged solution was diluted to 50 mL with pure water and filtered on 0.22 $\mu$m mixed cellulose ester membrane filters and the resultant graphene film transferred onto the gold mirror as described in the supplementary material to \cite{Wu-Rinzler-transparent-conductive-nanotube-films}. All steps in the transfer process are compatible with standard oxide and fluoride optical coatings as well as metal surfaces. Hence the graphene film could be used as a coating for a wide variety of optical components.

The Raman spectra of both the graphene film and the graphite flake starting material is displayed below in figure \ref{raman}. The relatively high D/G mode intensity ratio displayed by the graphene film is indicative of the introduction of edge defects brought about through sonication processing \cite{Lotya-Coleman-surfactant-stabilised-graphene-dispersion} and observed in other dispersion preparations which also involve sonication \cite{De-Coleman-surfactant-graphene-flake-film}. The shape of the 2D mode of the graphene film indicates the exfoliation of graphite flakes into few layer graphene \cite{Ferrari-Geim-graphene-raman-spectrum}, as required for efficient saturable absorption.

\begin{figure}[h]
\includegraphics*[width=3.1in]{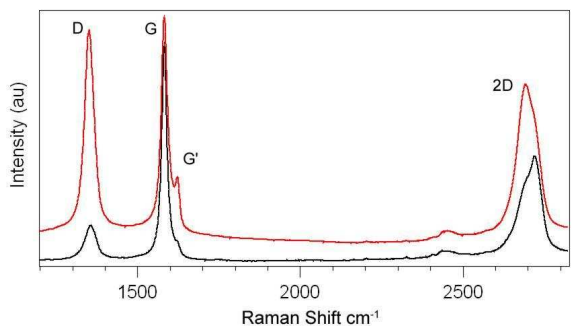}
\caption{Raman spectra showing the D, G and 2D modes of the graphene film on the gold mirror (red) and the graphite flake starting material (black). The laser excitation wavelength was 514 nm.}
\label{raman}
\end{figure}

\begin{figure}[htbp]
\includegraphics*[width=3.1in]{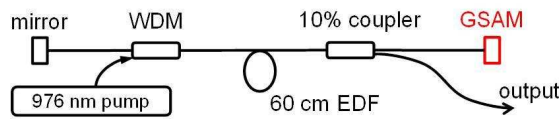}
\caption{Schematic of the FG-SAM mode-locked fiber laser. EDF = erbium-doped fiber, WDM = wavelength-division multiplexer.}
\label{schem}
\end{figure}

\begin{figure}[htbp]
\includegraphics*[width=3.1in]{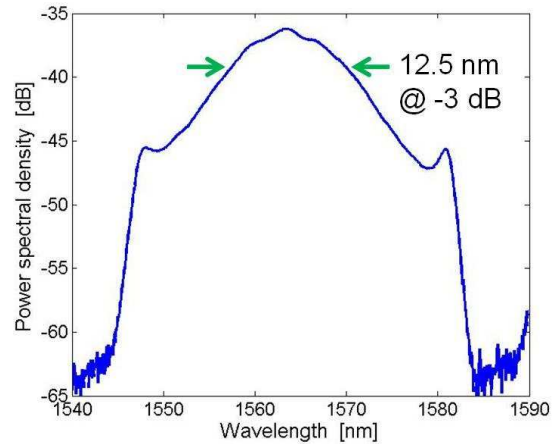}
\caption{Optical spectrum of the laser, showing typical soliton behavior. The optical bandwidth is 12.5 nm FWHM.}
\label{optspec}
\end{figure}

The FG-SAM was used to mode-lock an erbium-doped fiber laser, producing 190 fs pulses at a repetition rate of 42.8 MHz. The laser employed a standard linear-cavity design and a schematic is shown in Figure \ref{schem}. The lengths of standard single-mode fiber and erbium-doped gain fiber used in the laser were adjusted to give near-zero round-trip group velocity dispersion. At a pump power of 55 mW, mode-locked operation was initiated by sharply tapping one of the laser components. The laser bandwidth depended on adjustment of the polarization controller. Figure \ref{optspec} shows the optimized optical spectrum, with full-width at half-maximum (FWHM) of 12.5 nm. The spectrum is typical of soliton mode-locked lasers, with clearly visible soliton sidebands. The average laser power was 0.4 mW from each coupler port, corresponding to a circulating pulse energy of 90 pJ in the laser. For these pulse parameters, we calculate that the pulses circulating in the laser have soliton number $\sim 1$. At higher pump powers, we observed multiple pulse trains circulating in the laser. However, unlike in harmonically mode-locked fiber lasers, the relative time delay between pulse trains drifted over timescales of seconds.

\begin{figure}[htbp]
\includegraphics*[width=3.1in]{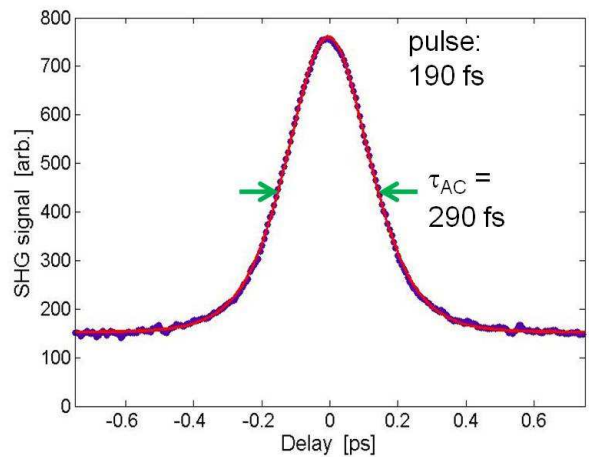}
\caption{Second-order autocorrelation of the compensated output pulses. Blue dots: data. Red line: sech$^2$ fit. The autocorrelation duration is 290 fs FWHM, corresponding to a pulse duration of 190 fs FWHM.}
\label{autocorr}
\end{figure}

The pulses were chirped at the output of the laser, but the chirp was compensated by propagation through $\sim 1$ m of single-mode fiber. The second-order autocorrelation trace of the resulting pulses is shown in Figure \ref{autocorr}, with autocorrelation FWHM of 290 fs. Assuming the $\mbox{sech}^2$ pulse shape typical of soliton mode-locked lasers, we obtain a pulse duration of 190 fs.

\begin{figure}[h]
\includegraphics*[width=3.1in]{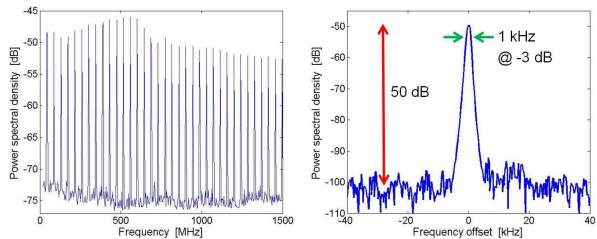}
\caption{Left: RF spectrum of laser photocurrent. Right: High-resolution RF spectrum of the fundamental repetition rate beatnote near 42.8 MHz.}
\label{rfpic}
\end{figure}

We measured the repetition rate characteristics of the laser by directing the laser output to a fast photodiode and analyzing the radiofrequency spectrum of the photocurrent. The repetition rate and its harmonics are seen as sharp spikes in the spectrum of Figure \ref{rfpic} (left). High signal-to-noise is maintained at high harmonics of the repetition rate, indicating low timing jitter in the output pulse train. The slight variation of harmonic amplitude is believed to be due to gain variations in the detection system. Figure \ref{rfpic} (right) shows a high-resolution spectrum of the  repetition rate beatnote near 42.8 MHz.

The FG-SAM exhibits large modulation depth and low saturated loss. The unsaturated loss of the FG-SAM was measured to be $55 \pm 5$\% by a low-power reflectivity measurement. The saturable absorption of the FG-SAM was quantified by measuring the laser slope efficiencies $\eta_\mathrm{ML,CW,bare}$ during mode-locking, in CW operation, and by using the bare gold substrate in place of the FG-SAM, which were found to be $(9.3 \pm 0.01) \times 10^{-3}$, $(8.2 \pm 0.1) \times 10^{-3}$, and $(11.2 \pm 0.2) \times 10^{-3}$, respectively. In each case, the slope efficiency is proportional to $1/(1+\delta_m/\delta_0)$, where $\delta_m$ is the loss in the mirror and $\delta_0$ is the remaining cavity loss \cite{Siegman-lasers-BOOK}. Hence the ratio of saturated to unsaturated loss is given by $(\eta_\mathrm{ML}^{-1} - \eta_\mathrm{bare}^{-1})/(\eta_\mathrm{CW}^{-1} - \eta_\mathrm{bare}^{-1})$, which evaluates to $0.57 \pm 0.03$ for our measurements. Since our FG-SAM is a reflective device, the single-pass saturated loss is actually $13 \pm 5$\%, a factor of 2 lower than previously reported for a flake-graphene saturable absorber \cite{Sun-Ferrari-graphene-Er-fiber, Martinez-Yamashita-optically-deposited-graphene-ML} and much lower than the 70\% loss obtained in previous work at sub-200-fs pulse duration \cite{Popa-Ferrari-graphene-Er-fiber-175fs}.

The saturation fluence of the absorber is estimated from the mode-locking threshold. Below 50 pJ circulating pulse energy, the laser dropped out of mode-locking. The $1/e^2$ mode-field diameter of SMF is specified by the manufacturer as 10.5 $\mu$m, so we estimate a saturation fluence of $\lesssim 100 \:\mu\mbox{J}/\mbox{cm}^2$, roughly in accord with other measurements on flake-graphene saturable absorbers \cite{Sun-Ferrari-graphene-Er-fiber, Popa-Ferrari-graphene-Er-fiber-175fs}.

The pulses from our FG-SAM mode-locked laser are below 200 fs and are within $10\%$ of the shortest pulse duration (175 fs) ever observed from a laser mode-locked with a graphene saturable absorber \cite{Popa-Ferrari-graphene-Er-fiber-175fs}. Our FG-SAM enables the use of graphene in a reflective saturable absorption geometry, which is generally preferred over the transmissive geometry in fiber and solid-state mode-locked lasers. Our FG-SAM exhibits low loss, enabling applications in relatively low-gain lasers. The graphene film used to make the FG-SAM can easily be deposited over areas of centimeters, making our device potentially useful in high-average-power mode-locked lasers that traditionally require carefully designed SESAMs \cite{Keller-compact-ultrafast-rev, Paschotta-Keller-ps-GHz-high-power-rev}.

This work was supported by the Australian Research Council under DP0773354 (Kielpinski) and by the School of Biomolecular and Physical Sciences and the Queensland Micro- and Nanotechnology Centre, Griffith University (Brown).

\end{document}